\documentclass[]{spie}  %>>> use for US letter paper
\usepackage{amsmath,amsfonts,amssymb}
\usepackage{graphicx}
\usepackage[colorlinks=true, allcolors=blue]{hyperref}
\usepackage{booktabs} % For prettier tables
\title{The Signal-to-Noise Ratio for Photon Counting After Photometric Corrections}

\author[a]{Kevin J. Ludwick}
\affil[a]{University of Alabama in Huntsville, 301 Sparkman Dr, Huntsville, AL 35899, USA}
\authorinfo{Further author information: (Send correspondence to K.J.L.)\\K.J.L.: E-mail: kevin.ludwick@uah.edu, Telephone: 1 256 824 2527}

\begin{document}
\vspace{1.5cm}
\vspace{1.5cm}
\vspace{1.5cm}
\maketitle
\newpage
\begin{abstract}
Photon counting is a mode of processing astronomical observations of low-signal targets that have been observed using an
electron-multiplying \textcolor{black}{charge-coupled device} (EMCCD).  In photon counting, the EMCCD amplifies the signal, and a thresholding technique effectively selects
for the signal electrons while drastically reducing relative noise sources.  Photometric corrections
have been developed which result in the extraction of a more accurate \textcolor{black}{estimate of the} signal of electrons, and the Nancy Grace Roman Telescope will utilize a
theoretical expression for the signal-to-noise ratio (SNR) given these corrections based on well-calibrated noise parameters to plan observations taken by its
coronagraph instrument.  I derive here analytic expressions for the SNR
for the method of photon counting, before and after these photometric corrections have been applied.
\end{abstract}

% Include a list of keywords after the abstract

\section{INTRODUCTION}
\label{sec:intro}  % \label{} allows reference to this section

Using a coronagraph to observe very low-signal exoplanets is very challenging.  The photon signal from
the planet must compete with \textcolor{black}{camera} noise sources such as clock-induced charge, dark current, and read noise,
where read noise is the most significant source and typically drastically swamps the signal.  The Nancy Grace
Roman Space Telescope (\textcolor{black}{Roman or} Roman Telescope for short) will employ an electron-multiplying \textcolor{black}{charge-coupled device} (EMCCD) for its coronagraph instrument (CGI)\cite{NGRST, Kasdin, Ilya, Rizzo}.  Utilizing
an EMCCD for observation allows the photo-electron signal to be amplified so that
the relative read noise is negligible, but other noise sources are also amplified.  If the photo-electron signal is small, obtaining the desired
\textcolor{black}{signal-to-noise ratio (SNR)} may not be possible when processing the EMCCD's output in the conventional, analog way.  \textcolor{black}{For example, a the CGI may want to image a dim exoplanet orbiting a relatively bright star.  Even with techniques for obtaining strong contrast between the star and the exoplanet, exposure time will be limited to prevent saturation of the detector from the star.  When the incident
photo-electrons on the EMCCD is on the order of $0.1$ electrons/pixel/frame or less\cite{Bijan}, photon counting is a
method that affords a higher SNR compared to the analog method.  Here, the SNR is defined as the photometric signal from the target of observation divided by the photon and camera noise contributions.  A typical SNR requirement is 5 to 7.  One can always increase the number of frames per observation to reach the SNR goal, and Roman can take thousands of frames per observation if necessary.}  

For a short exposure time for a series of frames taken by the EMCCD and a weak photon signal
from an observed target, most pixels in the detector will have zero electrons,
and occasionally there will be a few pixels with one electron each.  Very rarely will a pixel have two or more electrons.
So if a high multiplication (high gain) is applied, most pixels will read zero electrons while only a few will have some relatively high
number.  In photon counting, a threshold number of electrons is applied to each detector pixel.  The threshold is usually chosen to be several (e.g., five) \textcolor{black}{times the
standard deviation of the read noise probability density function}.  Other sources of noise, like clock-induced charge and dark current, %are included below this
\textcolor{black}{typically contribute orders of magnitude less than read noise.  The mean values of the clock-induced charge and dark current are} subtracted out if dark frames (frames taken in the absence of a target) are subtracted from the observation frames.
For a given pixel, if the number of post-gain counts is above the threshold, that pixel is designated as having one electron, and if
the number is below the threshold, the pixel is designated as having zero electrons.
The frames are then co-added to produce an image that has minimal input from read noise and \textcolor{black}{``}extra noise" inherent in the electron
multiplication process.  A short exposure time is important to minimize the chances \textcolor{black}{of damaging the EMCCD through saturation by the star the explanet is orbiting.  In addition, short exposure times can be helpful for minimizing the effect of cosmic rays and charge traps.  A} high gain is important   %to maximize the chances that only 0 or 1 photons interact with a given pixel, and a high gain is important
to maximize the chances that electrons are clearly below threshold (at 0) or clearly above.

For better accuracy, one should consider the inefficiency inherent in this thresholding.  The conversion of a photon to a photo-electron in a
detector pixel is a stochastic process, \textcolor{black}{as is} the multiplication in the gain register of the EMCCD.  \textcolor{black}{The stochastic nature of these processes leads to both overcounting and undercounting of electrons. Undercounting occurs when a pixel's number of post-gain counts is above the threshold (and is thus counted as 1 electron) when in fact there are actually more than one pre-gain electron for the pixel.  This undercount is known as ``coincidence loss" \cite{Bijan}.  Overcounting happens when a pixel's number of post-gain counts is below the threshold when in fact the pixel has a real electron.}  This effect is known as \textcolor{black}{``}threshold loss" \cite{Bijan}.  A third effect is leakage due to
detector effects.  For example, because the read noise is Gaussian in nature, thresholding five standard deviations above
the mean read noise will not completely cut out electrons due to read noise because of the tail of the distribution that
is above the threshold.  This is usually a small effect if the threshold is chosen to be high enough above the read noise mean \cite{Bijan}, and this leakage effect is ignored in this work\textcolor{black}{.  T}aking into account the photometric corrections of coincidence loss
and threshold loss gives a more accurate estimate of the mean number of electrons. %\textcolor{black}{, and thus the error in the observational estimate of the SNR is lower.}

%%%%%%
\textcolor{black}{Consider} an observation consisting of many frames of the same target.  We can model the readout for a single pixel in a single frame, after gain has been applied, as a random variate $q$:
\begin{equation}
\label{readout}
q = b + x + r,
\end{equation}
where $b$ is the voltage bias, $r$ is the contribution from read noise and is a variate of the normal distribution,  and $x$ is a variate of the Erlang distribution $P_e(x|g,n)$ \cite{Basden}.  This distribution is given by
\begin{equation}
\label{Erlang}
P_e(x | g,n) = \frac{x^{n-1} e^{-x/g}}{g^n (n-1)!},
\end{equation}
where $x$ is the number of electrons after multiplication given that $n \geq 1$ electrons entered the gain register at
a gain value of $g$.  This distribution is accurate for large values of gain, and this is the case with photon counting.  If $n$ electrons in a given pixel enter the gain register, the gain is defined by the
mean number of electrons after multiplication through the gain register, $g n$.

The conversion of a photon to a photo-electron as it interacts with a detector pixel (before the application of gain) is a Poisson process, and so is the generation of dark current
and clock-induced charge\textcolor{black}{.}  If the mean expected number of electrons per pixel 
\textcolor{black}{(photo-electrons as well as noise electrons)}
is $\lambda$, the number of electrons $n$ actually produced per pixel \textcolor{black}{(before the application of gain)} is given by the Poisson distribution function,
\begin{equation}
\label{Poisson}
P_p (n | \lambda) = \frac{\lambda^n e^{-\lambda}}{n!}.
\end{equation}
\textcolor{black}{The}
electron counts per pixel are then amplified as
the detector frame of pixels goes through the gain register.  Therefore, $n$ in Eq. (\ref{Erlang}) is a Poisson variate.  \textcolor{black}{The relevant photo-electron signal in the pixel resulting from the observation of the astrophysical source for time $t_{fr}$ is $s$.  The expected value of $s$ is $\langle s \rangle = \phi \eta t_{fr}$, where $\phi$ is the photon flux (photons/s), $t_{fr}$ is the exposure time for the frame, and $\eta$ is the pixel's
quantum efficiency (electrons/photon).    Along with this photo-electron signal, dark current $i_d$ and clock-induced charge $CIC$ contribute to the number of electrons in this pixel.   The Poisson variate $n$ can be written as the sum of the aforementioned Poisson variates:}
\begin{equation}
n = s + i_d t_{fr} + CIC.
\end{equation}

%The mean value of electrons in this pixel is found in practice by averaging over all frames.  
The expected mean value, $\lambda_{br}$, is given by 
\begin{equation}
\label{electrons}
 \lambda_{br} = \langle n \rangle = \langle s \rangle  + \langle i_d \rangle t_{fr} + \langle CIC \rangle,
\end{equation}
where the brackets indicate the \textcolor{black}{expectation value.} %, and we assume unbiased estimators here so that the theoretical mean is implied by the brackets, not simply an estimate of it.  (We simulate data for which we know the expected mean precisely and will discuss it later on.)  
\textcolor{black}{The expected value} of $q$ can therefore be written as 
\begin{equation}
\label{dk_electrons}
\langle q \rangle = \langle b \rangle + \langle x \rangle + \langle r \rangle = \langle b \rangle + g \lambda_{br},
\end{equation}
where the \textcolor{black}{expected} mean for the Erlang distribution \textcolor{black}{$g \lambda_{br}$} is used \textcolor{black}{for $\langle x \rangle$}, along with the fact that the \textcolor{black}{expected} read noise mean is $\langle r \rangle =0$.   
Other sources may contribute to the number of electrons, such as fixed-pattern noise and electron counts due to astrophysical background.  Assuming voltage bias is subtracted out and fixed-pattern noise and astrophysical background are corrected for separately, the measured value of the mean number of electrons for this pixel is obtained by averaging the pixel's value over all the frames and dividing by the gain.  Since this process of measuring the mean value for a pixel is is the same for any given pixel, Eq. (\ref{electrons}) will be treated as the mean number of electrons per pixel per frame.  

\iffalse
In an observation, the photo-electron signal from the observation target is $\phi_{ij} \eta_{ij} t_{fr}$, where $\phi_{ij}$ is the photon flux (photons/s) per pixel per frame (where $ij$ specifies the row and column indices for the pixel), $t_{fr}$ is the average exposure time per frame, and $\eta_{ij}$ is the pixel's
quantum efficiency (electrons/photon).  Along with this photo-electron signal, dark current $i_d$ and clock-induced charge ($CIC$) contribute to the mean number of electrons per frame for a given pixel (or, in other words, per pixel per frame).  The
For an analog observation with $N$ frames, the 
after averaging over the frames, the mean number of electrons per pixel per frame, $\lambda_{br}$, is given by
{\bf add in equation for $q = $ without the brackets, including read noise and FPN and bias.}
\begin{equation}
\label{electrons}
 \lambda_{br} = \langle \phi_{ij} \eta_{ij} t_{fr} + i_d t_{fr} + CIC \rangle.
\end{equation}
We ignore voltage bias in the detector and fixed-pattern noise as these can be subtracted out and corrected for separately. 
\fi

For the Roman Telescope's operations, \textcolor{black}{stars will be imaged iteratively in a process that will optimize the contrast for an orbiting expolanet\cite{Kasdin}}.  The mean photon flux \textcolor{black}{of a given target star will} be known very precisely, \textcolor{black}{as well as the telescope's throughput,} with negligible uncertainty.  Before any measurement of a target is done, the quantum efficiency, dark current, and clock-induced charge will be calibrated, and the calibration involves averaging over very many frames, many more than would be needed for a target measurement, and so the error in the knowledge of these parameters is also negligible and does not enter our idealized expressions for noise or SNR that follow.  

%A master dark frame with well-calibrated noise parameters (such as fixed-pattern noise, CIC, and dark current) will be subtracted from each observation frame.  So after this is done, the frames are then photon-counted (i.e., made subject to the threshold).  Since the thresholding is nominally supposed to collect all 1-count pixels (although not perfectly, due to coincidence loss and threshold loss), the mean number of electrons should roughly agree with Eq. (\ref{mean}).  After averaging over $N$ observation frames, the mean number of electrons present per pixel per frame is

If $N$ observation frames are taken (whether photon-counting or not), statistically there will be counts due to {\it noise} electrons.  So to accurately capture the profile of electrons due to noise, some number $N_2 \geq N$ of dark frames must be taken and subtracted from our observation.  For the dark frames, the \textcolor{black}{expected} mean number of electrons per pixel per frame is
\begin{equation}
\label{lambda_dark}
\lambda_{dk} = \langle i_d \rangle t_{fr} + \langle CIC \rangle.
\end{equation}
So if one averages the \textcolor{black}{``}bright" (target observation) frames and averages the \textcolor{black}{``}dark" frames and then subtracts the averaged darks from the averaged brights, one would have \textcolor{black}{an estimate of} the mean {\it signal} per pixel per frame.  \textcolor{black}{Theoretically, we expect that to be}
\begin{equation}
\label{brdk}
\lambda_{br} - \lambda_{dk} =  \langle s \rangle.
\end{equation}
The variance will be due to the sum of the variance of the brights and darks.  Therefore, we define the ideal SNR per pixel per frame to aim for as
\begin{equation}
\label{SNR_obs}
SNR = \frac{\lambda_{br} - \lambda_{dk} } { \sqrt{\sigma_{br}^2 +\sigma_{dk}^2} }.
\end{equation}
%In the equation above and hereafter, the notation $\sigma(y)$ is used to denote the standard deviation over the variates that have a mean of $y$.  
In other words, the goal is to get \textcolor{black}{an expression} for $\lambda_{br} - \lambda_{dk}$ for the numerator and the \textcolor{black}{standard deviation} from the corresponding variates whose mean provided the numerator.

\textcolor{black}{Given inputs of (good estimates of) photon flux, $\eta$, $\langle i_d \rangle$, and $\langle CIC \rangle$, one can calculate the observation parameters (exposure time $t_{fr}$, number of exposure frames $N$, and gain $g$) needed to achieve a certain SNR.}
\textcolor{black}{In order to do this calculation}, one must have an analytic theoretical expression for the SNR per pixel in terms of the noise parameters and photo-electron signal $\lambda_{br} - \lambda_{dk} = \langle s \rangle$. \textcolor{black}{\textcolor{black}{The} goal of this paper is to find that}  \textcolor{black}{expression.}
%%%%%

The Roman Telescope's CGI will use photon counting and employ photometric corrections so that the error budget requirements are met \cite{Roman2}.  Methods of signal extraction from photon-counted observations have been studied in the literature \cite{Lantz, Hu}\textcolor{black}{.}  Bijan Nemati (2020) introduced an algorithm that \textcolor{black}{estimates} the mean number of electrons for a given detector pixel to within the requisite 0.5\% accuracy with respect to the true mean \cite{Bijan}.  \textcolor{black}{The relationship between the mean electron rate and the number of counts out of the gain register is nonlinear, and an algorithm is required to correct for this nonlinearity.  For a given pixel in a multi-frame observation, the algorithm extracts a more accurate estimate of the mean electron rate from the post-gain counts \textcolor{black}{than} the result of simply averaging the post-gain counts for that pixel over the frames after thresholding. The algorithm iteratively solves for this more accurate estimate of the} true mean number of electron counts per pixel from the number of measured counts per pixel \textcolor{black}{and takes} into account coincidence loss and threshold loss as applied to the measured number of counts.
%this by going beyond merely accounting for coincidence loss and threshold loss by correcting the mean number of true photo-electron counts for a given pixel based on what the measured number of counts is.
\textcolor{black}{The algorithm requires a complex manipulation of the measured number of counts, and the standard deviation and measured estimate of the SNR of the electron counts change as a result.  The error in \textcolor{black}{the} estimate of the photo-electron signal \textcolor{black}{is} made very small due to the algorithm.}  %It turns out that the SNR also increases as a result, which means that
%less time is required for exposing the CGI to a star for observation while still achieving a target SNR.  
\textcolor{black}{F}or a given target star with known low-uncertainty input parameters such as photon flux and noise parameters of the CGI detector, the Roman Telescope's CGI software will \textcolor{black}{calculate the} observation parameters such as detector exposure time, number of frames of exposure, and gain so that \textcolor{black}{a chosen} SNR is achieved \textcolor{black}{while minimizing the total observation time. Certain constraints are also accounted for, including} detector saturation and constraints on the allowed ranges for exposure time, number of frames, and gain.  These allowed ranges come from detector limitations, considerations of cosmic rays and charge transfer inefficiency, and a limit to the photo-electron flux on the detector as appropriate for photon counting (around 0.1 electrons/pixel/frame, as mentioned earlier). 
\textcolor{black}{Therefore,} it is important for observation planning to have an accurate and analytic expression for the theoretical SNR in terms of the
noise parameters and photon flux.  In the following sections, the SNR is derived and presented, before and after \textcolor{black}{photometric} corrections.

\textcolor{black}{\section{MEAN NUMBER OF ELECTRONS VS MEAN NUMBER OF COUNTS }} %CORRECTIONS TO THE MEAN PHOTON RATE}

In photon counting, $N$ frames are taken, and the mean number of electrons per pixel over all
the frames (before \textcolor{black}{the application of gain}) is $\lambda N$.  \textcolor{black}{Remember} that only observations for which $\lambda$ is small (typically 0.1 electrons/pixel/frame or less) are being considered.  \textcolor{black}{A}fter multiplication, the pixels
that pass the threshold are each designated 
as one electron count, and those that do not pass the threshold are designated as 0 counts.  %\textcolor{black}{Let the random variable equal to the number of 1-designated counts per pixel per frame be $c$.}  
The effect of thresholding makes the \textcolor{black}{expected} mean number of 1-designated counts per pixel per frame \textcolor{black}{smaller than the mean number $\lambda$ of electrons per pixel per frame.   Nemati (2020) has a derivation for \textcolor{black}{the mean number of 1-designated counts per pixel per frame}, and his result is}
\begin{equation}
\label{mean_Nemati}
\lambda \epsilon_c \epsilon_{th},  % \langle c \rangle 
\end{equation}
where $\epsilon_c$ corrects for
coincidence loss and $\epsilon_{th}$ corrects for threshold loss.  \textcolor{black}{His coincidence loss factor comes from calculating an expected  number of non-zero pre-gain counts according to the Poisson distribution and normalizing it, and the factor is}\cite{Bijan}
\begin{equation}
\label{coincidence}
\epsilon_c = \frac{1 - e^{- \lambda}}{\lambda}\textcolor{black}{.}
\end{equation}
\textcolor{black}{His threshold loss factor comes from a process similar to what I show in the next section in Eqs. (\ref{P}) and (\ref{P_lines}) for the truncated post-gain probability distribution, except the pre-gain 0-count term is ignored in his derivation.  His threshold loss factor is}\cite{Bijan}
\begin{equation}
\label{threshold}
\epsilon_{th} = e^{-\tau/g} \left( 1 + \frac{\tau^2 \lambda^2 + 2 g \tau \lambda (3+\lambda)}{2g^2(6+3\lambda+\lambda^2)} \right)\textcolor{black}{,}
\end{equation}
\textcolor{black}{where $\tau$ is the threshold number of electrons.}
I will derive the probability distribution governing photon-counting, derive the expected mean from it, and revisit the above expression for the expected mean\textcolor{black}{, Eq. (\ref{mean_Nemati}),} in the next section.  \textcolor{black}{It turns out that Eq. (\ref{mean_Nemati}) is only approximately correct, but it is a very good approximation.  It agrees with what I derive from the probability distribution to third order in $\lambda$.}
\section{Derivation of Probability Distribution for Photon Counting}
% Take out derivation of Bijan's factors??  I'm basically doing a more accurate version here
For a given pixel in a given frame, the probability of getting $x$ electrons after multiplication given a mean number of pre-gain counts $\lambda$ is given by
\begin{equation}
\label{P}
P^0 (x | \lambda) = 
	\begin{cases} 
		& C(\lambda) \cdot  \sum_{i=1}^{\infty} P_e(x | g, i) P_p(i | \lambda), ~~ i>0 \\
		& C(\lambda) \cdot P_p(0 | \lambda), ~~ i=0,
\end{cases}
\end{equation}
where $C(\lambda)$ is a normalization factor.  There is little error in truncating this sum to 3 terms instead of an infinite number of terms since $i$, the number of pre-gain electrons, is rarely more than 1 for a given pixel for photon-counting conditions ($\lambda<<1$).  The Erlang distribution is not applicable in the $i=0$ case since there is no gain when  there are 0 electrons.  I call the distribution in which the sum is truncated to 3 terms $P_3^0(x | \lambda)$, with a normalization factor we will call $C_3(\lambda)$ (similar to what is done in Nemati (2020)).  The factor $C_3(\lambda)$ is obtained through normalization\footnote{\textcolor{black}{Technically, the distribution $P_3^0$ should be discrete, but the Erlang distribution only approximates the post-gain distribution, and it is normalized assuming a continuous distribution.  Therefore, we integrate in Eqs. (\ref{P_lines}) and (\ref{P_lines2}).}}:
\begin{equation}
\label{P_lines}
\begin{split}
&1=C_3(\lambda) \int^\infty_0 dx~ P^0_3(x|\lambda)  \\
& 1 = C_3(\lambda) \left[ e^{-\lambda} + \int^\infty_0 dx \sum_{i=1}^3 P_e(x|i) P_p(i|\lambda) \right] \\ 
&1 = C_3(\lambda) \left[ e^{-\lambda} + \int^\infty_0 dx ~ \lambda e^{-\lambda} \frac{e^{-x/g}}{g} \left(1+\frac{\lambda x}{2g} + \frac{\lambda^2 x^2}{12g^2} \right) \right] \\
&  C_3(\lambda) = \frac{6 e^{\lambda }}{\lambda  (\lambda  (\lambda +3)+6)+6}.
\end{split}
\end{equation}
As discussed earlier, if the number of post-gain counts in a pixel is above a threshold $\tau$ (say, \textcolor{black}{five} times the \textcolor{black}{standard deviation of the} read noise), that pixel is deemed as having \textcolor{black}{one} electron.  The probability $\epsilon_{th}'$  of passing the threshold $\tau$ is given by
\begin{equation}
\label{P_lines2}
\begin{split}
\epsilon_{th}' &= \int_{\tau}^\infty  dx ~P^0_3(x|\lambda) =  C_3(\lambda) \int_\tau^\infty dx~ \lambda e^{-\lambda} \frac{e^{-x/g}}{g} \left(1+\frac{\lambda x}{2g} + \frac{\lambda^2 x^2}{12g^2} \right)  \\
&= \frac{\lambda  e^{-\frac{\tau }{g}} \left(2 g^2 (\lambda  (\lambda +3)+6)+2 g \lambda  (\lambda +3) \tau +\lambda ^2 \tau ^2\right)}{2 g^2 (\lambda  (\lambda  (\lambda +3)+6)+6)}.
\end{split}
\end{equation}

\textcolor{black}{Let the random variable equal to the number of 1-designated counts per pixel for $N$ frames be $c$.}  \textcolor{black}{For $N=1$ frame, $c$ can either be 1 or 0.} Therefore, the probability $P_T$ of having $c$ electrons \textcolor{black}{in a pixel of a frame} after thresholding is given by a binomial distribution:
\begin{equation}
\label{thresh_pdf}
P_T(c) = 
\begin{cases}
	& 1- \epsilon_{th}', ~~ c=0 \\
	& \epsilon_{th}', ~~ c=1 \\
	& 0 ~~ \rm{otherwise}.
\end{cases}
\end{equation}
This probability distribution is normalized, so the mean expected value is given by 
\begin{equation}
\label{mean}
\langle c \rangle = \sum_0^1 c P_T(c) = \epsilon_{th}',
\end{equation}
and the standard deviation is given by
\begin{equation}
\label{std}
\sigma = \sqrt{\langle c^2 \rangle - \langle c \rangle^2} = \sqrt{\epsilon_{th}'(1-\epsilon_{th}')}.
\end{equation}

Compare Eq. (\ref{mean}) with the expression for the mean from Nemati's work \cite{Bijan} from Eq. (\ref{mean_Nemati}), $\lambda \epsilon_c \epsilon_{th}$.  For $\lambda<<1$, 
$\epsilon_{th}'$ to fourth order in $\lambda$ is 
\begin{equation}
\epsilon_{th}' \approx \lambda  e^{-\frac{\tau }{g}} - \frac{\lambda ^2 \left(e^{-\frac{\tau }{g}} (g-\tau )\right)}{2 g}   +\frac{\lambda ^3 e^{-\frac{\tau }{g}} \left(2 g^2-4 g \tau +\tau ^2\right)}{12 g^2}- \frac{\lambda ^4 \left(e^{-\frac{\tau }{g}} \left(g^2-g \tau +\tau ^2\right)\right)}{12 g^2},
\end{equation}
and, to fourth order in $\lambda$,
\begin{equation}
\label{Bijan_exp}
\lambda \epsilon_c \epsilon_{th} \approx \lambda  e^{-\frac{\tau }{g}} -\frac{\lambda ^2 \left(e^{-\frac{\tau }{g}} (g-\tau )\right)}{2 g} +\frac{\lambda ^3 e^{-\frac{\tau }{g}} \left(2 g^2-4 g \tau +\tau ^2\right)}{12 g^2} -\frac{\lambda ^4 \left(e^{-\frac{\tau }{g}} \left(g^2-2 g \tau +2 \tau ^2\right)\right)}{24 g^2} .
\end{equation}
\textcolor{black}{Nemati's expression for the mean and the more accurate one derived here} are very close, as there is exact agreement to third order in $\lambda$. \textcolor{black}{Nemati's expression is currently employed \textcolor{black}{for applying photometric corrections} in photon-counting frames for the Roman Telescope, so we calculate the SNR \textcolor{black}{which accounts for photometric corrections} later in this work according to his expression.  The degree of agreement is sufficient for accurate photon counting, which Nemati demonstrates in his work\cite{Bijan}.}
%The difference between these two expressions amounts to the fact that Nemati \cite{Bijan} applies the efficiency factors for coincidence loss $\epsilon_c$ and thresholding loss $\epsilon_{th}$ to the Poisson mean $\lambda$ to get an estimated mean that works very well.  

Eq. (\ref{thresh_pdf}) is valid for a single frame.  For two frames, the distribution is the convolution $P_T * P_T$:
\begin{equation}
(P_T * P_T)(c) = \sum_{k=0}^\infty P_T(k) P_T(c-k) = (1-\epsilon_{th}') P_T(c) + \epsilon_{th}' P_T(c-1)\textcolor{black}{,} 
\end{equation}
where $c$ for $N=2$ can be 0, 1, or 2.  
For $N$ frames, the distribution is the convolution of $P_T$ with itself $N$ times, which is
\begin{equation}
\textcolor{black}{P_T^{* N}(c) =} \sum_{k=0}^{N-1} {N-1 \choose k} (1- \epsilon_{th}')^{N-1-k} ~\epsilon_{th}'^k P_T(c-k).
 \end{equation}
The distribution is already normalized by construction.  Only the $k=c$ and $k=c-1$ terms survive in the sum, so one can write more simply the expression for the distribution for $N$ frames:
\begin{equation}
\label{thresh_N_pdf}
P_T^{* N}(c) = (1-\epsilon_{th}')^{N-c} ~\epsilon_{th}'^c \left[ { N-1 \choose c} + { N-1 \choose c-1} \right].
\end{equation}
The mean value for the distribution for $N$ frames can be calculated and is, not surprisingly, $N$ times the mean for one frame,
\begin{equation}
\label{mean_N}
\langle c \rangle_N = N \epsilon_{th}'\textcolor{black}{.}
\end{equation}
\textcolor{black}{T}he calculated standard deviation is $\sqrt{N}$ times the standard deviation for one frame:
\begin{equation}
\label{sigma}
\sigma_N = \sqrt{\langle c^2 \rangle_N - \langle c \rangle_N^2} = \sqrt{N \epsilon_{th}'(1-\epsilon_{th}')}.
\end{equation}
\textcolor{black}{As one might expect,} the mean number of counts for $N$ frames used in Nemati's work, $N \lambda \epsilon_c \epsilon_{th}$, agrees with the exact  expression $N \epsilon_{th}'$ to third order in $\lambda$\textcolor{black}{, just as there was agreement in the $N=1$ case.}  

\textcolor{black}{Note that if one averages $N$ observed frames, the expected number of counts would be}  \textcolor{black}{
\begin{equation}
\label{mean_avg}
\langle\frac{ c }{N} \rangle_N = \epsilon_{th}'.
\end{equation}
The variance of a function $f(x)$ of one variable is given by 
\begin{equation}
\label{error_prop}
\sigma(f(x))^2 = \left( \frac{\partial f}{\partial x}(\langle x \rangle) \right)^2 \sigma(x)^2.
\end{equation}
Applying this, the corresponding standard deviation for the average of $N$ frames} is given by 
\begin{equation}
\label{sigma_avg}
\textcolor{black}{\sigma(\frac{c}{N}) = \sqrt{\frac{1}{N^2} \sigma_N^2} = \sqrt{ \epsilon_{th}'(1-\epsilon_{th}')/N}.}
\end{equation}

\textcolor{black}{In practice, the SNR goal for Roman is determined through an error-budget analysis, and using calibrated detector noise parameters and a reliable flux estimate for a target star as inputs, the exposure time, EM gain, and number of frames needed to achieve the SNR goal while minimizing the total integration time are calculated.  In other words, the SNR of a measurement is not independently ``measured".  In order to obtain a measured SNR independent of calibrated input parameters, one needs an ensemble of measurements of $N$ bright frames and $N_2$ dark frames for a given pixel; an ensemble allows the mean and variance to be calculated for a given pixel, which the measured SNR requires\footnote{\textcolor{black}{One could take the mean and variance for a given pixel over $N$ bright frames and $N_2$ dark frames.  However, this would only consider the measured variates of the distribution for one frame (i.e., Eq. (\ref{thresh_pdf})) and would not take advantage of the statistics of multiple frames.  The mean and variance would be independent of the number of frames (see Eqs. (\ref{mean}) and (\ref{std})), whereas using the number of counts from multiple frames as the measured variate boosts the SNR by a factor on the order of the square root of the number of frames (i.e., the ratio of the Eqs. (\ref{mean_avg}) and (\ref{sigma_avg})).}}.  In practice, only one such measurement would be made in observation.  However, I can simulate an ensemble of measurements given the input parameters, agnostically measure the SNR for a given pixel, and compare the result to what the theoretical SNR expression gives.  Additionally, I can also compare the measured signal to the true signal based on the input parameters.  I refer to the multiple simulated measurements as ``trials" in the proceeding sections.}

\textcolor{black}{The next two sections calculate the theoretically expected SNR and verify its validity using \textcolor{black}{the estimated SNR from simulated data}.  Section \ref{SNR Before Photometric Corrections} does not correct the simulated data with Nemati’s algorithm, and this section also verifies the derived probability distribution against the simulated data with a $\chi^2$ test.  Section \ref{SNR AFTER PHOTOMETRIC CORRECTIONS}  calculates the theoretically expected SNR given the application of Nemati's algorithm and verifies it using \textcolor{black}{the estimated SNR from simulated data}, where the data have been run through Nemati’s algorithm  in this section.}

%We truncate to $n=3$ as we discussed earlier, and the only difference between this and $P_n(x | \lambda)$ from Eq. (\ref{distribution}) is that we include the $x=0$ case. 

\section{SNR Before Photometric Corrections}
\label{SNR Before Photometric Corrections}
%After averaging over $N_{fr}$ frames, t

For an observation, the goal is to most accurately \textcolor{black}{estimate} the mean signal per pixel per frame, which is $\lambda_{br} - \lambda_{dk}$.  If $N$ observation frames are photon-counted (i.e., summed, subjected to the threshold, and averaged over the $N$ bright frames), statistically there will be 1-designated counts due to noise electrons.  So to counter this, $N_2 \geq N$ dark frames must be photon-counted and then subtracted from the observation.  However, all that is available for a given pixel location is the number of 1-designated counts over the frames, which is not the same thing as the number of electron counts because of coincidence loss and threshold loss.  However, one could naively treat the number of 1-designated counts per pixel %(random variable $N_{br}$ for the bright frames and random variable $N_{dk}$ for the dark frames) 
divided by the respective number of frames as the number of electrons per pixel per frame \textcolor{black}{(i.e., apply no photometric corrections via Nemati's algorithm)}.  Let $N_{br}$ be the total number of 1-designated counts over all the bright frames for a given pixel, and let $N_{dk}$ be the total number of 1-designated counts over all the dark frames for the pixel.  \textcolor{black}{These are both observed variates of Eq. (\ref{thresh_N_pdf}).  The measured average number of counts for that pixel over all bright frames is \textcolor{black}{$ N_{br}/N$}, and the analogous measured average for the dark frames is \textcolor{black}{$ N_{dk}/N_2$}.} %Let $N_{br}$ be what is theoretically expected given $\lambda_{br}$, and let $N_{dk}$ be what is theoretically expected given $\lambda_{dk}$, and we will denote these counts as measured in an observation as $\hat{N_{br}}$ and $\hat{N_{dk}}$. 

\textcolor{black}{As discussed in the previous section, an ensemble of measurements (i.e., the ``trials"), each with $N$ bright and $N_2$ dark frames, will be simulated in order to measure the SNR.  The measured estimate of the SNR over an ensemble of measurements is }
\begin{equation}
\label{SNR_unc_obs_sin}
SNR_{unc, obs} = \textcolor{black}{\frac{\bar{N}_{br}/N - \bar{N}_{dk}/N_2}{\sqrt{\sigma(N_{br}/N)^2 +\sigma(N_{dk}/N_2)^2}},}
%\frac{\textcolor{black}{ \bar{c}_{br,obs} } -  \textcolor{black}{\bar{ c}_{dk,obs}} } { \sqrt{\sigma(c_{br,obs})^2 +\sigma(c_{dk,obs})^2} },
\end{equation}
where the subscript $obs$ stands for \textcolor{black}{``}observational"\textcolor{black}{,}  $unc$ stands for \textcolor{black}{``}uncorrected"\textcolor{black}{, and $~\bar{}~$ over a variable denotes the ensemble mean of the variable}.  
%, and $sin$ stands for "single".  
(The \textcolor{black}{``}corrected" version will be discussed later.)  \textcolor{black}{Here, \textcolor{black}{$\sigma(N_{br}/N)$ and $\sigma(N_{dk}/N_2)$} simply come from computing the standard deviation for the pixel under consideration over the \textcolor{black}{trials}.}

Eq. (\ref{mean_avg}) implies \textcolor{black}{that we expect $\bar{N}_{br}/N$ to be $ \epsilon_{th}'(\lambda_{br})$ and $\bar{ N}_{dk}/N$ to be $ \epsilon_{th}'(\lambda_{dk})$.  Eq. (\ref{sigma_avg}) implies that we expect $\sigma(N_{br}/N)^2$ to be $ \epsilon_{th}'(\lambda_{br})(1-\epsilon_{th}'(\lambda_{br}))/N$ and $\sigma(N_{dk}/N_2)^2$ to be $\epsilon_{th}'(\lambda_{dk})(1-\epsilon_{th}'(\lambda_{dk}))/N_2$.}  Therefore, the corresponding theoretical expression, assuming knowledge of $\lambda_{br}$ and $\lambda_{dk}$, is 
\begin{equation}
\label{SNR_unc_th}
\textcolor{black}{SNR_{unc, th} = \frac{ \epsilon_{th}'(\lambda_{br}) - \epsilon_{th}'(\lambda_{dk}) } { \sqrt{ \epsilon_{th}'(\lambda_{br}) (1-\epsilon_{th}'(\lambda_{br}))/N + \epsilon_{th}'(\lambda_{dk}) (1-\epsilon_{th}'(\lambda_{dk}))/N_2}}\textcolor{black}{,}}
\end{equation}
\textcolor{black}{where $th$ stands for ``theoretical".}

If one wanted to test Eq. (\ref{SNR_unc_obs_sin}) with simulated data to see if it aligns with theoretical expectations \textcolor{black}{in Eq. (\ref{SNR_unc_th})}, one could \textcolor{black}{first} compare the numerator of Eq. (\ref{SNR_unc_th}), the signal, to the numerator of Eq. (\ref{SNR_unc_obs_sin}) with its uncertainty\textcolor{black}{, and then do the same for the denominators}.  \textcolor{black}{Consider a\textcolor{black}{n ensemble} of a random variable $b$ measured $M$ times with a\textcolor{black}{n ensemble} mean of $\bar{b}$ and a standard deviation of $\sigma_b$.  For uncertainty of the \textcolor{black}{ensemble} mean ($\delta \bar{b}$)}, the standard \textcolor{black}{error} of the mean is used:
\begin{equation}
\label{mean_unc_general}
\textcolor{black}{\delta \bar{b} = \sigma_b/\sqrt{M}}.
\end{equation}
The uncertainty \textcolor{black}{$\delta \bar{C}_{obs}$ of the numerator of Eq. (\ref{SNR_unc_obs_sin}), denoted by \textcolor{black}{$\bar{C}_{obs} \equiv \bar{N}_{br}/N - \bar{N}_{dk}/N_2$},} %$\mu_{unc,obs} \equiv \langle c_{br,obs} \rangle -  \langle c_{dk,obs} \rangle$ 
is found by applying Eq. (\ref{mean_unc_general}) separately to the brights and darks and adding in quadrature\textcolor{black}{.  For $M$ trials in an ensemble of measurements, }
\begin{equation}
\label{mean_unc}
\textcolor{black}{\delta \bar{C}_{obs} } = \textcolor{black}{ \sqrt{ \frac{\sigma(N_{br}/N)^2}{M} + \frac{\sigma(N_{dk}/N_2)^2}{M}}}.
\end{equation}
\textcolor{black}{T}he denominator of Eq. (\ref{SNR_unc_th}) would be compared to the denominator of Eq. (\ref{SNR_unc_obs_sin}) with its uncertainty.   The fractional uncertainty of the standard deviation, $1/\sqrt{2(M-1)}$, is utilized for the uncertainty of the standard deviation, $\delta \sigma$, so that\footnote{This formulation for the uncertainty in the standard deviation assumes an approximately normal distribution, and this approximation is valid if all possible values of variates (0 through $M$) fal\textcolor{black}{l} within \textcolor{black}{three} standard deviations of the mean for the distribution, and this is true if $M > 9(1-\epsilon_{th}')/\epsilon_{th}'$ and $M > 9 \epsilon_{th}'/(1-\epsilon_{th}')$.  This implies that \textcolor{black}{ $M>94.9$, and $M=500$ trials is used in the simulation}.}
\begin{equation}
\label{std_unc_general}
\textcolor{black}{\delta \sigma_b = \frac{\sigma_b}{\sqrt{2(M-1)}}.}
\end{equation}
\textcolor{black}{In the case of the simulated data}, the uncertainty of the standard deviation is found by applying Eq. (\ref{std_unc_general}) separately to brights and darks and adding in quadrature.  \textcolor{black}{Denoting the denominator of Eq. (\ref{SNR_unc_th}) as $\sigma_{obs}\equiv \sqrt{\sigma(N_{br}/N)^2 +\sigma(N_{dk}/N_2)^2}$, one finds}
\begin{equation}
\label{std_unc}
\textcolor{black}{\delta \sigma_{obs}} = \textcolor{black}{ \sqrt{ \frac{\sigma(N_{br}/N)^2}{2(M-1)} + \frac{\sigma(N_{dk}/N_2)^2}{2(M-1)} }}.
\end{equation}

%This can be done for a single measurement of $N$ bright frames and $N_2$ dark frames; the comparison can also be done over several trials of the observation, using the average of \textcolor{black}{$\bar{c}_{obs}$, $\delta \bar{c}_{obs}$, $\sigma_{obs}$, and $\delta \sigma_{unc,obs}$ over the trials}.  The latter is done in the demonstration below.  

%could only get a single value for the observational SNR for a single observation.  The SNR value would vary from observation to observation, though.  So for a reliable comparison, we should do multiple trials to get an observational SNR with an uncertainty, with each observation consisting of $N$ bright frames and $N$ dark frames.  In that sense, for a given pixel, $N_{br}$ and $N_{dk}$ can also be thought of as variates of the probability distribution from Eq. (\ref{thresh_N_pdf}), and we can use the mean number of electrons per pixel for each of those variates ($\langle N_{br} \rangle$ and $\langle N_{dk} \rangle$), along with their standard deviations ($\sigma(N_{br})$ and $\sigma(N_{dk})$), from Eqs. (\ref{mean_N}) and (\ref{sigma}).  The observational SNR over all these trials per pixel per frame is then 
%\begin{equation}
%\label{SNR_unc_obs}
%SNR_{unc, obs} = \frac{\langle N_{br} \rangle_N/N - \langle N_{dk} \rangle_N/N } { \sqrt{\sigma_N(N_{br} )^2/N +\sigma_N(N_{dk} )^2/N} }.
%\end{equation}

%In fact, Eq. (\ref{SNR_unc_th}) is the theoretical SNR per pixel per frame that follows from both Eq. (\ref{SNR_unc_obs_sin}) and Eq. (\ref{SNR_unc_obs}).  

A simulated photon flux map consisting of 50x50 pixels each with a value of 1 photon/s was simulated and run through the Python module
{\tt emccd\textunderscore detect}\footnote{Publicly available here: https://github.com/wfirst-cgi/emccd\_detect.}, which is a EMCCD detector simulator.  It created \textcolor{black}{$N=600$} detector frames with a quantum efficiency for each pixel of $0.9$ electrons/photon (or $e^{-}$/photon), and each frame had an exposure time of \textcolor{black}{$0.1$} s.  Typical values
for read noise (standard deviation of 100 $e^{-}$/pixel/frame), dark current (mean value of $8.33*10^{-4}$ $e^{-}$/s/pixel/frame), and CIC (mean value of 0.01 $e^{-}$/pixel/frame) are included, and $\lambda_{br}$, as given by Eq. (\ref{electrons}),
   is \textcolor{black}{$0.10 e^{-}$}.  For darks, \textcolor{black}{$N_2=800$} frames with the same noise parameters were also simulated, and $\lambda_{dk}$, as given by Eq. (\ref{dk_electrons}), is $0.01 e^{-}$.  I then photon-count these frames (threshold each of the frames and then sum them, separately for brights and darks) using the Python module {\tt PhotonCount}\footnote{Publicly available here:  https://github.com/wfirst-cgi/PhotonCount.}
with a gain of $5000$ and a threshold of $500 e^{-}$. %which outputs the mean number of counts for each pixel.  
This process is done for 500 trials (500 sets of $N=\textcolor{black}{600}$ brights and $N_2=\textcolor{black}{800}$ darks) for each pixel.  \textcolor{black}{Each pixel is under identical conditions and independent from the others, so I average the signal and noise and their uncertainties over the 50x50 pixels.  Of course, averaging over multiple pixels is not necessary for the analysis, and one pixel is all that is needed; averaging over the 50x50 pixels simply ensures robustness of the results.}  The script to do this analysis
is {\tt noise\textunderscore script.py}\footnote{https://github.com/wfirst-cgi/PhotonCount/blob/main/noise\_script.py}, publicly available in the {\tt PhotonCount} module, and one can adjust the input parameters as desired.  

I will demonstrate the accuracy of Eq. (\ref{SNR_unc_th})\textcolor{black}{,} but first, I examine the probability distribution for $N$ bright frames, Eq. (\ref{thresh_N_pdf}) (and the probability distribution is also \textcolor{black}{applicable to} dark frames).  Fig.\ 1 was created by {\tt noise\_script.py}, and it shows a histogram of $N_{br}$, the pixels' values for the summed bright frames, along with the probability distribution, Eq. (\ref{thresh_N_pdf}).  Since there are 500 trials for each of the 50x50 pixels, there is a total of 1.25e6 entries in the histogram.  The histogram counts were normalized for comparison with the probability distribution.  The $\chi^2$ value for the fit of the data to the probability distribution is \textcolor{black}{$4.71 \times 10^{-3}$}, and the critical $\chi^2$ value in this case is \textcolor{black}{$111$}, which indicates the data and the probability distribution are not statistically distinct.  The $p$ value is exactly 1.0 to machine precision, which indicates that the null hypothesis should be accepted (i.e., the data follows the distribution very well).  The darks were created in the same way, so they also follow the probability distribution well.  

\begin{figure} [ht]
   \begin{center}
   %\begin{tabular}{c} %% tabular useful for creating an array of images
   \includegraphics[height=7cm]{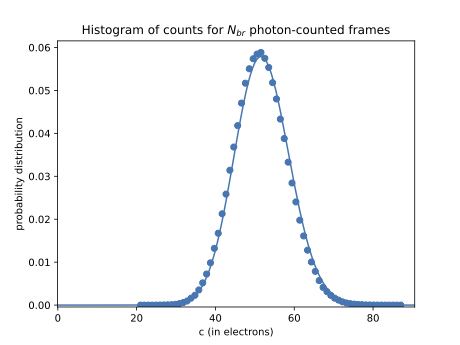}
   %\end{tabular}
   \end{center}
   \caption[]
%>>>> use \label inside caption to get Fig. number with \ref{}
   { \label{prob_hist}
  This plot shows a histogram of $N_{br}$, the pixels' electron count values (horizontal axis), generated with {\tt PhotonCount} (represented by the dots), along with the theoretical probability distribution, Eq. (\ref{thresh_N_pdf}) (represented by the curve).  An input photon flux map
   was fed through {\tt emccd\textunderscore detect} to obtain simulated detector frames, \textcolor{black}{600} of them, and they were then photon-counted using {\tt PhotonCount}.
   This process was done 500 times for each of the 50x50 pixels for the sake of robust statistics, and a $\chi^2$ fit was performed.  Noise sources are included.  See the text for specific values for input parameters and details on the fit.  The mean value is $N \epsilon_{th}'(\lambda_{br}) = \textcolor{black}{52.0} e^{-}$.  The $\chi^2$ value for the fit is $\textcolor{black}{4.71 \times 10^{-3}}$, and the critical $\chi^2$ value in this case is \textcolor{black}{$111$}, which indicates the data and the probability distribution are not statistically distinct.  The $p$ value is exactly 1.0 to machine precision, which indicates that the null hypothesis should be accepted (i.e., the data follows the distribution well). }
   \end{figure}
   
I now examine the uncorrected observational SNR per pixel for $N=\textcolor{black}{600}$ bright frames and $N_2 = \textcolor{black}{800}$ dark frames.  For the same simulation parameters described above, Eq. (\ref{SNR_unc_obs_sin}) is computed for each \textcolor{black}{of the 500 trials} for each of the 50x50 pixels.
%\footnote{We mask out any pixels for which there is a standard deviation of 0 for a given trial.  This would either lead to a signal of 0 if no electron counts were present in any of the $N$ frames or to an infinite SNR if the counts for each of the $N$ frames were 1 (which is highly unlikely).  This can be avoided if $N$ is increased and/or if the exposure time is increased.}. 
The uncorrected observational signal (using the numerator of Eq. (\ref{SNR_unc_obs_sin}) and Eq. (\ref{mean_unc})) is averaged over these trials and all pixels, and the result is %$0.1684 \pm 0.0038 e^{-}$.  
$\textcolor{black}{0.0775 \pm 0.0005 e^{-}}$.  The uncorrected theoretical signal (numerator of Eq. (\ref{SNR_unc_th})) is $\textcolor{black}{0.0775 e^{-}}$ and is in agreement.  The uncorrected observational noise (using the denominator of Eq. (\ref{SNR_unc_obs_sin}) and Eq. (\ref{std_unc})) is $\textcolor{black}{0.0119 \pm 0.0004 e^{-}}$, and the uncorrected theoretical noise (denominator of Eq. (\ref{SNR_unc_th})) is $\textcolor{black}{0.0120 e^{-}}$ and agrees.  Using these uncertainties, the inferred observational SNR range is $\textcolor{black}{(6.25, 6.75)}$, and the average signal over the average noise is $\textcolor{black}{6.49}$.  The theoretical SNR is \textcolor{black}{6.48} and is in agreement with this range.

I will now examine how to get an accurate estimate of the quantity of interest, $\lambda_{br} - \lambda_{dk}$\textcolor{black}{,} based on what is observationally available.  I will then find the \textcolor{black}{``}corrected" SNR \textcolor{black}{(i.e., uses Nemati's algorithm to apply photometric corrections)}.

%%%%%%%%%%%%%%%%%%%%%%%

\section{SNR AFTER PHOTOMETRIC CORRECTIONS}
\label{SNR AFTER PHOTOMETRIC CORRECTIONS}

%Now consider $N_{br}$ and $N_{dk}$ as variates of Eq. (\ref{thresh_N_pdf}).  
If $N$ bright frames and $N_2$ dark frames are taken for a single observation, the \textcolor{black}{expected} mean number of 1-designated counts per pixel is given by Eq. (\ref{mean}), and, as stated earlier, the expression is equal to the analogous expression from Nemati's\cite{Bijan} work to third order in $\lambda_{br}$:
\begin{equation}
\label{Nbr}
%N_{br}/N = \epsilon_{th}'(\lambda_{br}) \approx  \lambda_{br} \epsilon_c(\lambda_{br}) \epsilon_{th}(\lambda_{br}).
\textcolor{black}{\epsilon_{th}'}(\lambda_{br}) \approx  \lambda_{br} \epsilon_c(\lambda_{br}) \epsilon_{th}(\lambda_{br})\textcolor{black}{.}
\end{equation}
\textcolor{black}{As mentioned before, Nemati's expression will be used since it is used in {\tt PhotonCount} and by the Roman Telescope software.  The result is accurate, too, as will be demonstrated.  An} analogous expression relating \textcolor{black}{$\epsilon_{th}'(\lambda_{dk}) $} and $\lambda_{dk}$ can be written as well.  The goal is to get a more accurate approximation to the desired SNR in Eq. (\ref{SNR_obs}).  \textcolor{black}{I first consider the application of Nemati's algorithm for photometric corrections to data, and then I compute the theoretical SNR, which effectively \textcolor{black}{expresses the desired mean electron rate in terms of observable variates}.}  

\textcolor{black}{For an observation, the observable variate after photon-counting is the average number of 1-designated counts, but the desired quantity is the observational estimate of the mean number of electrons.  Since \textcolor{black}{$N_{br}$} is the best observational estimate of Eq. (\ref{Nbr}), Nemati's algorithm essentially equates these two things.  For the sake of precision, when these are ``equated", let $\lambda_{br}$ be replaced with the observational best estimate that results from solving the equation, denoted as $L_{br}$: 
\begin{equation}
\label{Nbr2}
\textcolor{black}{N_{br}} =  L_{br} \epsilon_c(L_{br}) \epsilon_{th}(L_{br}).
\end{equation}
Also, one can do the analogous process for $N_2$ dark frames to extract $L_{dk}$ so that one can determine the observational best estimate for $\lambda_{br} - \lambda_{dk}$, which is $L_{br} - L_{dk}$.  I will go through the analysis for just the brights in what follows, but the process is exactly the same for the darks.
}
  
%Eq. (\ref{Nbr_N}) can be solved for $\lambda_{br}$ to correct for the photometric effects of coincidence loss and threshold loss\textcolor{black}{, which make the mean number of 1-designated counts differ from the mean number of electrons.}  
%\textcolor{black}{As mentioned before, the} expression from Nemati's work will be used below for solving for $\lambda_{br}$ and the SNR since the expression is used in {\tt PhotonCount} and by the Roman Telescope software.  The result is accurate, too, as will be demonstrated.  

In Eq. (\ref{Nbr2}), $\epsilon_c$ and $\epsilon_{th}$ are functions of \textcolor{black}{$L_{br}$}, and there is
no exact analytic solution for \textcolor{black}{$L_{br}$}, and we would like an analytic expression for expressing the corrected observational SNR.  An analytic expression is also useful for the speed of an algorithm that finds \textcolor{black}{$L_{br}$} for every single pixel.  One can get an approximate analytic expression for \textcolor{black}{$L_{br}$}
using Newton's method, which is an iterative method which gets closer to the solution with each iteration.  Nemati
came up with this solution and tested it\cite{Bijan}, and two iterations of Newton's method is sufficient
for a very accurate solution.  \textcolor{black}{The presentation of Newton's method as applied to this problem follows Nemati's treatment.}

Newton's method requires an initial guess \textcolor{black}{$L_{guess}$} as an input\textcolor{black}{.  The} initial guess used is the solution to Eq. (\ref{Nbr2}) \textcolor{black}{that results from truncating} $\textcolor{black}{L_{br}} \epsilon_c \epsilon_{th}$ to first order in \textcolor{black}{$L_{br}$.  The expression to first order can be written down easily by using} Eq. (\ref{Bijan_exp}).  The result is
\begin{equation}
\label{lambda_guess}
%\langle N_{br} \rangle_N \approx N \lambda_{br} e^{\frac{-\tau}{g}} ~\rightarrow~ \lambda_{guess} = - \ln \left( 1 - \frac{ \langle N_{br} \rangle_N/N}{e^{-\tau/g}} \right).
\textcolor{black}{N_{br} \approx N L_{br} e^{\frac{-\tau}{g}} ~\rightarrow~ L_{guess} = - \ln \left( 1 - \frac{  N_{br}/N}{e^{-\tau/g}} \right).}
\end{equation}
Newton's method estimates the next iteration in terms of the previous:
\begin{equation}
\textcolor{black}{L_{n+1} = L_{n} - \frac{f(L_n)}{f'(L_n)}},
\end{equation}
where \textcolor{black}{$L_n$} is the $n$th iteration for the root of the equation
\begin{equation}
\label{f}
\textcolor{black}{f(L_{br}) = \textcolor{black}{N}_{br} \epsilon_c(L_{br}) \epsilon_{th}(L_{br}) - N_{br}.}
\end{equation}
\textcolor{black}{Using Eqs. (\ref{coincidence}) and (\ref{threshold}) in} Eq. (\ref{f}) results in \textcolor{black}{$f(L_{br})$}, and \textcolor{black}{$f'(L_{br})$ is
\begin{equation}
\begin{split}
f'(L_{br}) & = \frac{e^{-\tau/g - L_{br}} N}{(2 g^2 (6 + 3 L_{br} + L_{br}^2)^2} ~(2 g^2 (6 + 3 L_{br} + L_{br}^2)^2 + \tau^2 L_{br} (-12 + 3 L_{br} + 3 L_{br}^2 + L_{br}^3 + 3 e^{L_{br}} (4 + L_{br})) \\
& + 2 g \tau (-18 + 6 L_{br} + 15 L_{br}^2 + 6 L_{br}^3 + L_{br}^4 + 6 e^{L_{br}} (3 + 2 L_{br}))).
\end{split}
\end{equation}}
Starting with \textcolor{black}{$L_{guess}$} and iterating twice results in a long analytic but unwieldy expression for
\textcolor{black}{$L_2$}.  To get \textcolor{black}{a} manageable expression\textcolor{black}{,} I had to expand
\textcolor{black}{$L_2$} for \textcolor{black}{$ N_{br} /N \ll 1$}, and I found that truncating to third order was sufficient.  This expansion is valid since
\textcolor{black}{$L_{br} \ll 1$} for effective photon counting (as described previously), and from the rightmost part of Eq. (\ref{Nbr2}), it follows that \textcolor{black}{$ N_{br}/N < L_{br}$} since the factors $\epsilon_c$ and $\epsilon_{th}$ must be less than 1 (see Eqs. (\ref{coincidence}) and (\ref{threshold})).  To third order, \textcolor{black}{$L_2$} for the brights is
\begin{equation}
\label{L_2}
\textcolor{black}{L_{2, br} = \frac{e^{\tau/g}  N_{br} }{N} + \frac{e^{2\tau/g}  N_{br}^2 (g-\tau)}{2 g N^2} + \frac{e^{3\tau/g}   N_{br}^3 (4g^2-8g \tau+5\tau^2)}{12g^2 N^3}.}
\end{equation}
One can see that \textcolor{black}{$L_{2, br}$} is smaller than 1 even in the minimal case of $N = 1$ since $\tau/g < 1$ (typically $\sim 1/10$, as it is in the case of the aforementioned simulations) and \textcolor{black}{$ N_{br}$} would only be 1 at most in that case. 
%and this approximation to third order works
%reasonably well even in this case (as one can see using {\tt noise\textunderscore script.py}).  
Of course, for accurate photon counting, $N$ will certainly be bigger than 1.

The \textcolor{black}{``}corrected" observational estimate of the SNR per pixel per frame is obtained by considering the \textcolor{black}{mean and variance} over the multiple trials \textcolor{black}{(as discussed earlier)} of observations:
\begin{equation}
\label{SNR_obs_corr}
%SNR_{corr,obs} =  \frac{\lambda_{2, br}  - \lambda_{2, dk} } { \sqrt{\sigma_{N,br}^2 +\sigma_{N_2,dk}^2} }
%SNR_{corr} =  \frac{\langle L_{2, br} \rangle_N - \langle L_{2, dk} \rangle_{N_2} } { \sqrt{\sigma_N(L_{2, br} )^2 +\sigma_{N_2}(L_{2, dk} )^2} }
\textcolor{black}{SNR_{corr,obs} =  \frac{\bar{L}_{2, br}  - \bar{L}_{2, dk} } { \sqrt{\sigma(L_{2, br} )^2 +\sigma(L_{2, dk} )^2} }}
\end{equation}
where $corr$ stands for \textcolor{black}{``}corrected"\textcolor{black}{.}  %This equation will stand for both the observational corrected and theoretical corrected SNR.  
For a given pixel, one would calculate this observationally by finding $N_{br}$ (and $N_{dk}$) for each trial and then computing $L_{2, br}$ (and $L_{2, dk}$) with Eq. (\ref{L_2}), and then one would find \textcolor{black}{$\bar{L}_{2,br}$ (and $\bar{L}_{2,dk}$)} by averaging over all the trials.  The standard deviation of $L_{2,br}$ and $L_{2, dk}$ over all the trials would provide the denominator of Eq. (\ref{SNR_obs_corr}).  %Eq. (\ref{SNR_obs_corr}) is used for both observational and theoretical SNR; the theoretical version would be too long to express concisely, but the parts needed for the expression above are expressed theoretically in the equations that follow.

 I now turn to the theoretical corrected SNR per pixel, which assumes true knowledge of $\lambda_{br}$ and $\lambda_{dk}$.  \textcolor{black}{It is the theoretically expected version of Eq. (\ref{SNR_obs_corr}):
 \begin{equation}
SNR_{corr,th} =  \frac{\langle {L}_{2, br} \rangle_N  - \langle{L}_{2, dk}\rangle_{N_2} } { \sqrt{\sigma(L_{2, br} )_N^2 +\sigma(L_{2, dk} )_{N_2}^2} },
\end{equation}
where, examining Eq. (\ref{L_2}) and using $c_{br}$ of Eq. (\ref{thresh_N_pdf}), 
\begin{equation}
\langle L_{2, br} \rangle_N = \frac{e^{\tau/g} \langle c_{br}\rangle_N }{N} + \frac{e^{2\tau/g}  \langle c_{br}^2 \rangle_N (g-\tau)}{2 g N^2} + \frac{e^{3\tau/g}  \langle c_{br}^3 \rangle_N (4g^2-8g \tau+5\tau^2)}{12g^2 N^3},
\end{equation}
and a similar relation would hold for $\langle L_{2, dk} \rangle_{N_2}$.  The standard deviations in the denominator are found using Eq. (\ref{sigma}).}

 \textcolor{black}{For the numerator, $\langle c_{br} \rangle_N$, $\langle c_{br}^2 \rangle_N$, and $\langle c_{br}^3 \rangle_N$} need to be computed.  Eq. (\ref{mean_N}) already provides \textcolor{black}{$\langle c_{br} \rangle_N$}.  One can compute (using, for example, {\it Mathematica}) \textcolor{black}{$\langle c_{br}^2 \rangle_N$ and $\langle c_{br}^3 \rangle_N$}:
\begin{equation}
\label{mean_N2}
\textcolor{black}{\langle c_{br}^2 \rangle_N }= \sum_0^{N} c^2 P^{*N}_T(c) = N \epsilon_{th}' \left( \lambda_{br}) (1+(N-1) \epsilon_{th}'(\lambda_{br}) \right),
\end{equation}
\begin{align}
\label{mean_N3}
%\langle N_{br}^3 \rangle = \sum_0^{N} x^3 P^{*N}_T(x)= \frac{\epsilon_{th}'(\lambda_{br}) \left((1-\epsilon_{th}'(\lambda_{br}))^n \left(-\, _4F_3\left(2,2,2,1-n;1,1,1;1+\frac{1}{\epsilon_{th}'(\lambda_{br})-1}\right)\right)-(n-1) (\epsilon_{th}'(\lambda_{br})-1)^2 ((n-2) \epsilon_{th}'(\lambda_{br}) ((n-3) \epsilon_{th}'(\lambda_{br})+3)+1)\right)}{\epsilon_{th}'(\lambda_{br})-1}.
\textcolor{black}{\langle c_{br}^3 \rangle_N} = \sum_0^{N} c^3 P^{*N}_T(c)= & \frac{\epsilon_{th}'(\lambda_{br})}{1- \epsilon_{th}'(\lambda_{br})} \bigg((1-\epsilon_{th}'(\lambda_{br}))^N \, _4F_3\left(2,2,2,1-N;~1,1,1;~1+\frac{1}{\epsilon_{th}'(\lambda_{br})-1}\right) + \nonumber \\ 
& (N-1) (1- \epsilon_{th}'(\lambda_{br}))^2 ((N-2) \epsilon_{th}'(\lambda_{br}) ((N-3) \epsilon_{th}'(\lambda_{br})+3)+1)\bigg),
\end{align}
where $\, _4F_3$ is a generalized hypergeometric function.  

The \textcolor{black}{theoretically expected standard deviation of $L_{2, br}$ is needed for the denominator of the theoretical corrected SNR, and Eq. (\ref{sigma}) along with error propagation is used to find it}.  
Applying \textcolor{black}{Eq. (\ref{error_prop})}, it follows that
\begin{equation}
\label{sigma2_lambda2}
\textcolor{black}{\sigma(L_{2,br})_N^2 = \left(  \frac{e^{\tau/g}}{N} + 2  \frac{e^{2\tau/g} (g-\tau)}{2 g N^2}  \langle c_{br} \rangle_N + 3 \frac{e^{3\tau/g} (4g^2 - 8g \tau +5 \tau^2)}{12 g^2 N^3} \langle c_{br} \rangle_N^2    \right)^2 \sigma(c_{br})_N^2,}
\end{equation} 
\textcolor{black}{and the analogous expression for $\sigma(L_{2,dk})_{N_2}^2$ is also obtained in this way.}

Using the previous equations, one has expressions for all the pieces needed \textcolor{black}{for the theoretical corrected SNR.}

\textcolor{black}{I now compare the theoretical corrected SNR to the observational corrected SNR by analyzing simulated data}.  I simulated 500 trials for each of the 50x50 pixels under the same conditions as described in Sect. \ref{SNR Before Photometric Corrections} and found the observational corrected SNR as described below Eq. (\ref{SNR_obs_corr}).  The numerator (signal) and denominator (noise) of the SNR are examined separately.  Eqs. (\ref{mean_unc_general}) and (\ref{std_unc_general}) provide the uncertainty of the signal and noise respectively, where $M$ in those equations is \textcolor{black}{again} equal to 500, the number of trials per pixel for both brights and darks, and $\textcolor{black}{b \equiv  L_{2, br}  - L_{2, dk}}$.  Then, instead of examining a single instance of the signal and noise that was obtained over 500 trials, I average \textcolor{black}{$\bar{b}$, $\delta \bar{b}$, $\sigma_{b}$, and $\delta \sigma_{b}$} over the 50x50 pixels.  \textcolor{black}{The pixels are under identical conditions and are independent, and I averaged over them just as I did in the ``uncorrected" case.}

For the observational corrected signal, I find \textcolor{black}{$0.0901 \pm 0.0006 e^{-}$}, and the theoretical corrected signal is \textcolor{black}{$0.0901 e^{-}$} and agrees.  For the observational corrected noise, I find \textcolor{black}{$0.0144 \pm 0.0005 e^{-}$}, and the theoretical corrected noise is \textcolor{black}{$0.0144 e^{-}$} and agrees.  Using these uncertainties, the inferred observational SNR range is \textcolor{black}{$(6.03, 6.52)$}, and the average signal over the average noise is \textcolor{black}{6.27}.  The theoretical SNR is \textcolor{black}{6.27} and is in agreement.  Note that $\lambda_{br} - \lambda_{dk} = \textcolor{black}{0.1e^- - 0.01e^- = 0.09e^-}$, which \textcolor{black}{disagrees with the observational uncorrected signal but} agrees with the observational corrected signal\textcolor{black}{.  This agreement with the corrected observational signal} demonstrates that Nemati's expression for the mean (approximation used in Eq. (\ref{Nbr2})) is good.

The corrected SNR, using \textcolor{black}{my example simulated} parameters, is \textcolor{black}{a little less than the uncorrected SNR, and this is true for both the observational and theoretical cases.  The decrease is not substantial, though, and the uncorrected case has a higher SNR for the ``wrong" signal (mean number of 1-designated counts instead of mean number of electrons from the target).  I have demonstrated that the corrected expressions more accurately probe the target.}  As one might naturally expect, as the number of frames (bright or dark) increases, the theoretical corrected signal approaches the Poisson mean ($\lambda_{br}$ or $\lambda_{dk}$), and the theoretical corrected noise approaches 0.  \textcolor{black}{A reasonable SNR is achieved, and increasing the number of observation frames makes the signal more accurate and the SNR higher.}  

\textcolor{black}{Table 1} summarizes the results for the \textcolor{black}{``}uncorrected" and \textcolor{black}{``}corrected" analysis.

\begin{table}[h!]
  \begin{center}
    \label{table1}
    \begin{tabular}{c | c | c} %{l | S | r}
      \toprule % <-- Toprule here
      {}   & \textbf{Observational} & \textbf{Theoretical}\\
      \hline
      \midrule % <-- Midrule here
      \textbf{uncorrected signal} & $\textcolor{black}{0.0775 \pm 0.0005e^-}$ & \textcolor{black}{$0.0775e^-$}\\
      \textbf{uncorrected noise} & $\textcolor{black}{0.0119 \pm 0.0004e^-}$ & \textcolor{black}{$0.0120e^-$}\\
      \textbf{uncorrected SNR} & \textcolor{black}{$(6.25, 6.75)$}, average: \textcolor{black}{6.49} & \textcolor{black}{6.48} \\
      \hline
      \textbf{corrected signal} & \textcolor{black}{ $0.0901 \pm 0.0006e^-$} & \textcolor{black}{$0.0901e^-$}\\
      \textbf{corrected noise} & \textcolor{black}{$0.0144 \pm 0.0005e^-$} & \textcolor{black}{$0.0144e^-$}\\
      \textbf{corrected SNR} & \textcolor{black}{$(6.03, 6.52)$}, average: \textcolor{black}{6.27} & \textcolor{black}{6.27} \\
      \bottomrule % <-- Bottomrule here
    \end{tabular}
    \caption{Summary of the \textcolor{black}{``}uncorrected" and \textcolor{black}{``}corrected" simulations and theoretical expectations.  Simulation parameters used:  $\lambda_{br} = \textcolor{black}{0.1}e^{-}$, $\lambda_{dk} = 0.01e^{-}$, $N=\textcolor{black}{600}$, $N_2 = \textcolor{black}{800}$, $g=5000$, and $\tau = 500e^{-}$. The same simulation parameters were applied for each frame's 50x50 pixels, and the number of trials of photon-counting for these $N$ bright frames and $N_2$ dark frames was 500.}
  \end{center}
\end{table}

\section{Conclusion}

\textcolor{black}{In summary, I have derived theoretical expressions for the SNR per pixel for a photon-counted stack of frames, with and without photometric corrections, along with the binomial probability distribution from which the expressions followed.  These expressions are useful for coronograph missions that utilize photon counting for observations with low signal flux.  The SNR expression which accounts for photometric corrections will be used by the Roman Telescope's CGI to determine observational parameters.  The error budget demands a low level of uncertainty in the knowledge of the mean number of photo-electron counts for a given pixel, and the photometric corrections come from an algorithm developed by Bijan Nemati \cite{Bijan} which meets this requirement and is employed in {\tt PhotonCount}.  I show with simulations that the observed signal without these corrections does not agree with the actual signal, even though the uncorrected SNR is slightly higher than the corrected SNR for the same number of frames per observation.  For a given set of low-error input parameters for the target star and noise properties of the CGI's detector,
Roman's software will determine the \textcolor{black}{detector} exposure time, number of frames, and gain \textcolor{black}{to achieve a chosen SNR} \textcolor{black}{while minimizing the integration time}.} 

%The Roman Telescope will use the SNR expression which accounts for photometric corrections for determination of the minimum observation time which still achieves its SNR goal.  I verify the SNR expressions and the probability distribution with simulations.}  

%I have derived the theoretical mean and standard deviation of the number of photo-electron counts per pixel per frame for a photon-counted stack of frames.  
%The process of designating photo-electron signals per pixel per frame above a certain threshold follows a binomial distribution \textcolor{black}{which I derived}, and the mean and standard deviation follow straightforwardly.  %I then show how applying photometric corrections based on the measured number of 1-designated counts for a given pixel using Newton's method corrects the mean and standard deviation, and I calculate the theoretical expressions for these.  
I provide {\tt noise\textunderscore script.py}, available to the public\footnote{{https://github.com/wfirst-cgi/PhotonCount/blob/main/noise\_script.py}}, which utilizes the publicly available Python modules {\tt emccd\textunderscore detect} and {\tt PhotonCount} (links in a previous footnote) to simulate photon-counted detector frames.  The script
also demonstrates how the simulations agree with the theoretical expressions, and the results of the script are provided in this paper.  
%The error in the knowledge of the true SNR is minimized with our corrected SNR, and it is also larger than the expected SNR without applying corrections, which affords lower exposure times for reaching a target SNR.

\section{Code, Data, and Materials Availability}
The script that does the analysis in this paper, {\tt noise\textunderscore script.py}, is freely available at the public GitHub repository:  

{https://github.com/wfirst-cgi/PhotonCount/blob/main/noise\_script.py}

\noindent The script is in {\tt PhotonCount} repository and utilizes the modules {\tt PhotonCount} and {\tt emccd\textunderscore detect}, which must be installed before using the script.  They are publicly available:

\noindent  {\tt emccd\textunderscore detect} freely available here:
 
{https://github.com/wfirst-cgi/emccd\_detect}

\noindent  {\tt PhotonCount} freely available here:  

{https://github.com/wfirst-cgi/PhotonCount}

\noindent Simple installation instructions are included.  All code is in Python.

\acknowledgments % equivalent to \section*{ACKNOWLEDGMENTS}

This work was done under contract with the Jet Propulsion Laboratory, California Institute of Technology.
The author would like to thank Dr. Bijan Nemati for useful discussions.

\noindent A previous version of this paper appeared on arXiv: \\
Kevin J. Ludwick, {\tt arXiv:2203.16501 [astro-ph.IM]}, 2022.  

% References
\bibliography{report} % bibliography data in report.bib

\begin{thebibliography}{1}

\bibitem{NGRST}
Mennesson, B., Debes, J., Douglas, E., Nemati, B., Stark, C., Kasdin, J.,
  Macintosh, B., Turnbull, M., Rizzo, M., Roberge, A., Zimmerman, N., Cahoy,
  K., Krist, J., Bailey, V., Trauger, J., Rhodes, J., Moustakas, L., Frerking,
  M., Zhao, F., Poberezhskiy, I., and Demers, R., ``{WFIRST} coronagraph
  instrument: A major step in the exploration of sun-like planetary systems via
  direct imaging,'' {\em Proceedings of SPIE - The International Society for
  Optical Engineering}~{\bf 10698,} (2018).

\bibitem{Kasdin}
Kasdin, N.~J., Bailey, V.~P., Mennesson, B., Zellem, R.~T., Ygouf, M., Rhodes,
  J., Luchik, T., Zhao, F., Riggs, A. J.~E., Seo, B.-J., Krist, J., Kern, B.,
  Tang, H., Nemati, B., Groff, T.~D., Zimmerman, N., Macintosh, B., Turnbull,
  M., Debes, J., Douglas, E.~S., and Lupu, R.~E., ``{The Nancy Grace Roman
  Space Telescope Coronagraph Instrument (CGI) technology demonstration},'' in
  [{\em {Space Telescopes and Instrumentation 2020: Optical, Infrared, and
  Millimeter Wave}}{\nolinebreak\hspace{0.1em}]},  Lystrup, M., Perrin, M.~D.,
  Batalha, N., Siegler, N., and Tong, E.~C., eds.,  {\bf 11443},  114431U,
  International Society for Optics and Photonics, SPIE (2020).

\bibitem{Ilya}
{Poberezhskiy}, I., {Luchik}, T., {Zhao}, F., {Frerking}, M., {Basinger}, S.,
  {Cady}, E., {Colavita}, M.~M., {Creager}, B., {Fathpour}, N., {Goullioud},
  R., {Groff}, T., {Morrissey}, P., {Kempenaar}, J., {Kern}, B., {Koch}, T.,
  {Krist}, J., {Mok}, F., {Muliere}, D., {Nemati}, B., {Riggs}, A.~J., {Seo},
  B.-J., {Shi}, F., {Shreckengost}, B., {Steeves}, J., and {Tang}, H., ``{Roman
  space telescope coronagraph: engineering design and operating concept},'' in
  [{\em Space Telescopes and Instrumentation 2020: Optical, Infrared, and
  Millimeter Wave}{\nolinebreak\hspace{0.1em}]},  {Lystrup}, M. and {Perrin},
  M.~D., eds., {\em Society of Photo-Optical Instrumentation Engineers (SPIE)
  Conference Series} {\bf 11443},  114431V (2021).

\bibitem{Rizzo}
{Rizzo}, M.~J., {Groff}, T.~D., {Zimmermann}, N.~T., {Gong}, Q., {Mandell},
  A.~M., {Saxena}, P., {McElwain}, M.~W., {Roberge}, A., {Krist}, J., {Riggs},
  A.~J.~E., {Cady}, E.~J., {Mejia Prada}, C., {Brandt}, T., {Douglas}, E., and
  {Cahoy}, K., ``{Simulating the WFIRST coronagraph integral field
  spectrograph},'' in [{\em Society of Photo-Optical Instrumentation Engineers
  (SPIE) Conference Series}{\nolinebreak\hspace{0.1em}]},  {Shaklan}, S., ed.,
  {\em Society of Photo-Optical Instrumentation Engineers (SPIE) Conference
  Series} {\bf 10400},  104000B (2017).

\bibitem{Bijan}
Nemati, B., ``{Photon counting and precision photometry for the Roman Space
  Telescope Coronagraph},'' in [{\em Space Telescopes and Instrumentation 2020:
  Optical, Infrared, and Millimeter Wave}{\nolinebreak\hspace{0.1em}]},
  Lystrup, M., Perrin, M.~D., Batalha, N., Siegler, N., and Tong, E.~C., eds.,
  {\bf 11443},  884 -- 895, International Society for Optics and Photonics,
  SPIE (2020).

\bibitem{Basden}
Basden, A.~G., Haniff, C.~A., and Mackay, C.~D., ``Photon counting strategies
  with low-light-level {CCD}s,'' {\em Monthly Notices of the Royal Astronomical
  Society}~{\bf 991},  985--991 (2003).

\bibitem{Roman2}
Harding, L.~K., Demers, R.~T., Hoenk, M., Peddada, P., Nemati, B., Cherng, M.,
  Michaels, D., Neat, L.~S., Loc, A., Bush, N., Hall, D., Murray, N., Gow, J.,
  Burgon, R., Holland, A., Reinheimer, A., Jorden, P.~R., and Jordan, D.,
  ``Technology advancement of the {CCD}201-20 {EMCCD} for the {WFIRST}
  coronagraph instrument: sensor characterization and radiation damage,'' {\em
  Journal of Astronomical Telescopes, Instruments, and Systems}~{\bf 2},
  011007 (2015).

\bibitem{Lantz}
Lantz, E., Blanchet, J.-L., Furfaro, L., and Devaux, F., ``{Multi-imaging and
  Bayesian estimation for photon counting with EMCCDs},'' {\em Monthly Notices
  of the Royal Astronomical Society}~{\bf 386}(4),  2262--2270 (2008).

\bibitem{Hu}
{Hu}, M.~M., {Sun}, H., {Harness}, A., and {Kasdin}, N.~J., ``{Bernoulli
  generalized likelihood ratio test for signal detection from photon counting
  images},'' {\em Journal of Astronomical Telescopes, Instruments, and
  Systems}~{\bf 7},  028006 (2021).

\end{thebibliography}
\bibliographystyle{spiebib} % makes bibtex use spiebib.bst

\section*{BIOGRAPHIES}

Kevin Ludwick is a Principal Research Scientist at the University of Alabama at Huntsville, in the Center for Applied Optics.  He does research in image processing, optics calibration, and theoretical cosmology.  He earned his Ph.D. in Physics at the University of North Carolina and was a Pirrung Postdoctoral Fellow at the University of Virginia.  He was then a professor at LaGrange College.  He has served on the executive committee for the APS FECS.

\end{document}